\documentclass{article}
\usepackage[english]{babel}
\usepackage[utf8]{inputenc}
\usepackage{johd}
\usepackage{listings}
\usepackage{xcolor}
\title{Rennes Metropolis Air Quality Dataset: \\ Microsensor-Based Measurements}
\author{François Bodin$^{a}$$^{*}$, Laurent Morin$^{a}$, Matthieu Adafaly$^{a}$, Marius Garénaux$^{a}$\\
        \small $^{a}$Irisa, University of Rennes, Rennes, France \\\\
        \small $^{*}$Corresponding author: François, Bodin; \tt{bodin@irisa.fr} \\
}
\date{May 2025}

\begin{document}
\maketitle
\begin{abstract} 
\noindent The dataset gathers particulate matter (PM\(_{2.5}\)) measurements collected using fixed and mobile micro-sensors (AlphaSense OPC-N3) in the Rennes Metropolitan area from April 1 to December 31, 2020.

\end{abstract}
\noindent\\

\noindent\authorroles{\\
François Bodin: Conceptualization, Supervision, Investigation\\
Laurent Morin: Investigation, Methodology, Software, Data curation, Visualization\\
Matthieu Adafaly : Extraction of smaller dataset, Data publication \\
Marius Garénaux : Extraction of smaller dataset, Data paper writing, Data publication \\
}\\

\section{Overview}

\paragraph{Repository location} The full dataset is available via the Rennes Metropolis Open Data platform: \url{https://rudi.rennesmetropole.fr/catalogue/detail/af03e9a4-2faa-4f7b-8fcf-cb55f892b964/donnees-brutes-mesures-de-la-qualite-de-l-air-dans-la-metropole-rennaise-fr-a-l-aide-de-microcapteurs-mobiles}. To get only measures of a specific month, see Section \ref{sec:dataset_desc}.

This dataset has been produced by the AQMO project [\cite{aqmo_website}] funded by the Connecting Europe Facility (CEF). We present in this data paper both the raw dataset from the AQMO project, and an extraction of it, in tabular format, that can be used for data analysis (see Section \ref{sec:light} for details).

Air quality in urban areas and the impact of pollution on citizens' health have become major concerns over the past decades, leading public administrations to make significant investments in assessing this risk. Currently, air quality monitoring is conducted at the city scale using highly accurate but costly fixed measurement stations placed in a few strategic locations. The AQMO air pollution measurement campaign aims to enhance the current monitoring approach by deploying a mobile micro-sensor platform in the Rennes metropolitan area. This approach provides a more representative measurement scale for air pollution variability, capturing transient situations that fixed surveillance stations and simulations might miss. 

\begin{figure}[!ht]
    \centering
    \includegraphics[width=0.8\textwidth]{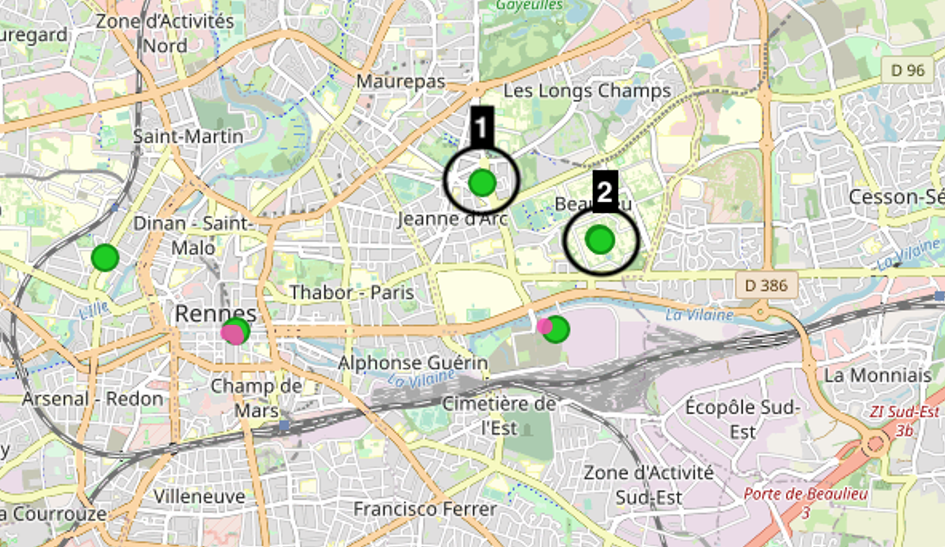}
    \caption{This map shows two fixed sensors: the first is located on Boulevard Charles Péguy, a road with traffic, while the second is on the University Campus, an area with little traffic.}
    \label{fig:fixed_sensors}
\end{figure}

Between April and December 2020, ten sensors measuring fine particulate matter (PM2.5) were gradually installed at two fixed locations (Figure \ref{fig:fixed_sensors}) and on eight metropolitan buses. The Method Department of Keolis selected these buses (and corresponding routes) to maximize spatial coverage.

\begin{figure}[!ht]
    \centering
    \includegraphics[width=0.8\textwidth]{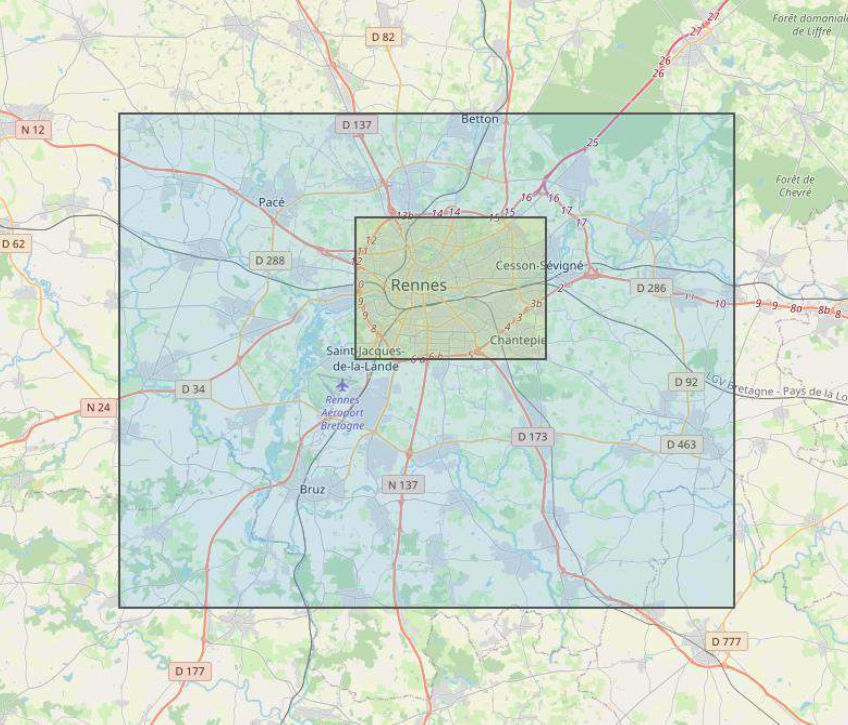}
    \caption{This map shows the geographical coverage of the 2020 measurement campaign. The inner square accounts for 96\% of the measurements.}
    \label{fig:coverage}
\end{figure}

As of this report, these sensors have generated a substantial amount of data, with over 1.2 million measurements collected over nearly 2,000 aggregated days. The total area covered is approximately 672 km², though most of the measurements were taken within a smaller zone of 60 km² (see Figure \ref{fig:coverage}).

\noindent The data set covers the period from April 1st to December 31th 2020. 

\noindent The main data fields include:
\begin{itemize}
    \item The PM\(_{2.5}\) value in µg/m³,
    \item The measurement timestamps,
    \item The GPS coordinates or path of the measurement if mobile.
\end{itemize}

\noindent More details on the AQMO measurement campaign can be found in [\cite{aqmo_d2_2}].

\section{Mobile Sensors based Data Capture}

The data are collected using a network of micro-sensors (OPC-N3 [\cite{alphasense_opc_n3}]). The dataset focuses on fine particulate matter (PM\(_{2.5}\)), which includes particles with a diameter of less than \(2.5\,\mu m\). 

Details on sensor characteristics, performance evaluation, and comparisons with regulatory air quality monitoring instruments are provided in AQMO Report D6.2 [\cite{aqmo_d6_2}].

\subsection{Sensors Positioning}

\begin{figure}[!ht]
    \centering
    \includegraphics[width=0.8\textwidth]{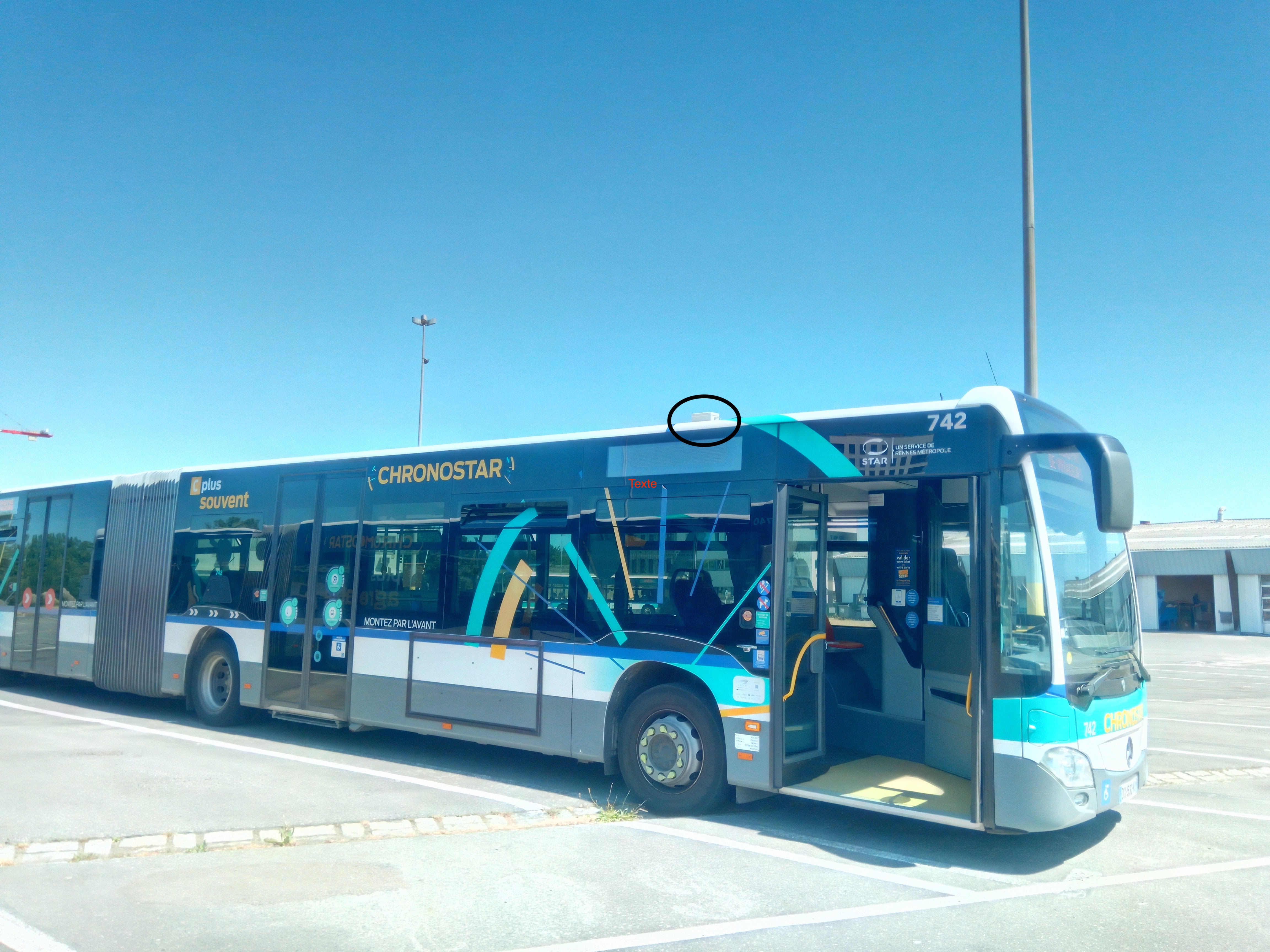}
    \caption{A STAR bus equipped with an AlphaSense OPC-N3 sensor. The sensor is mounted at the front of the bus, positioned sufficiently far from the exhaust to minimize contamination from emissions}
    \label{fig:aqmo_points}
\end{figure}

Eight mobile sensors were installed on the roofs of buses.  The sensor position is shown in Figure~\ref{fig:aqmo_points}. A more detailed description of the installation of the mobile sensor platform is available in AQMO report D6.1~[\cite{aqmo_d6_1}].

\section{Dataset Description}\label{sec:dataset_desc}

The full dataset is available on the RUDI portal : \url{https://rudi.rennesmetropole.fr/catalogue/detail/af03e9a4-2faa-4f7b-8fcf-cb55f892b964/donnees-brutes-mesures-de-la-qualite-de-l-air-dans-la-metropole-rennaise-a-l-aide-de-microcapteurs-mobiles}. Here are the specific download links for each month :

\begin{itemize}
    \item \href{https://opendata.fenix.rudi-univ-rennes1.fr/storage/download/42547d6a-736c-429e-972d-355a971707d1}{2020-04-01 to 2020-05-01},
    \item \href{https://opendata.fenix.rudi-univ-rennes1.fr/storage/download/d090f2d5-1c86-48bd-991b-46c1a76e2f0e}{2020-05-01 to 2020-06-01},
    \item \href{https://opendata.fenix.rudi-univ-rennes1.fr/storage/download/26a78e18-f093-4d66-9cba-d91a122759ad}{2020-06-01 to 2020-07-01},
    \item \href{https://opendata.fenix.rudi-univ-rennes1.fr/storage/download/3272a000-aaf9-447f-a065-ea17332cb766}{2020-07-01 to 2020-08-01},
    \item \href{https://opendata.fenix.rudi-univ-rennes1.fr/storage/download/22682787-348f-4b4b-9bcc-0bc37e373bec}{2020-08-01 to 2020-09-01},
    \item \href{https://opendata.fenix.rudi-univ-rennes1.fr/storage/download/2b2ae028-bbac-4b5b-8006-46f7debc1961}{2020-09-01 to 2020-10-01},
    \item \href{https://opendata.fenix.rudi-univ-rennes1.fr/storage/download/e90b23f1-5fe0-4d25-8d08-7bc88a06df3a}{2020-10-01 to 2020-11-01},
    \item \href{https://opendata.fenix.rudi-univ-rennes1.fr/storage/download/96d6ebd8-fd73-4a0e-97fa-059973508801}{2020-11-01 to 2020-12-01},
    \item \href{https://opendata.fenix.rudi-univ-rennes1.fr/storage/download/c4609776-3097-4e58-9a0b-157a79a767c0}{2020-12-01 to 2021-01-01}.
\end{itemize}

\paragraph{Repository name} RUDI (\url{https://rudi.rennesmetropole.fr/home})
\paragraph{Object name} Données brutes - Mesures de la Qualité de l'Air dans la Métropole Rennaise (FR) à l'Aide de Microcapteurs Mobiles 
\paragraph{Format names and versions} Compressed json (.json.zip)
\paragraph{Creation dates} (2020-04-01) to (2020-12-31) .
\paragraph{Dataset creators} 
\begin{itemize}
    \item François Bodin : Conceptualization, Supervision, Investigation, Data paper writing
    \item Laurent Morin : Investigation, Methodology, Software, Data curation, Visualization, Data paper writing
    \item Matthieu Adafaly : Extraction of smaller dataset, Data publication
    \item Marius Garénaux : Extraction of smaller dataset, Data paper writing, Data publication 
\end{itemize}
\paragraph{Language} English
\paragraph{License} Etalab Licence Ouverte 2.0 (\url{https://www.etalab.gouv.fr/licence-ouverte-open-licence/})
\paragraph{Publication date} 2025-05-19
\\

Sensor measurement entries are natively structured using the SenML format, which includes the measurement date and all associated sensor data. At any given time, the association between each sensor and a specific station is known. When sensor data is received, sensor entries are dynamically generated by combining all station location entries that match the measurement period. Each new location entry is represented as either a \texttt{"Point"} or a \texttt{"LineString"} in GeoJSON terminology. The detailed scheme can be accessed here : \url{https://data.aqmo.org/schemas/geojson}.

The resulting sensor entries include:
\begin{itemize}
    \item A GeoJSON geographical description using either a \texttt{"Point"} coordinate or a path constructed with a \texttt{"LineString"} set of coordinates,
    \item A fusion of the station properties,
    \item A list of SenML sensor data values stored as properties.
\end{itemize}

It is worth noting that issues persist with two mobile sensors that are not being properly recorded in the dataset. This data gap is due to incorrect installation in buses 701 and 915.

\section{A More Concise Dataset}\label{sec:light}

In addition to the raw dataset presented above, we propose a more concise version, in tabular format. This smaller dataset is an extraction of the raw one. It contains time series of particle matter measurements for each sensor; and geographical information.

\begin{figure}[!ht]
    \centering
    \includegraphics[width=1\textwidth]{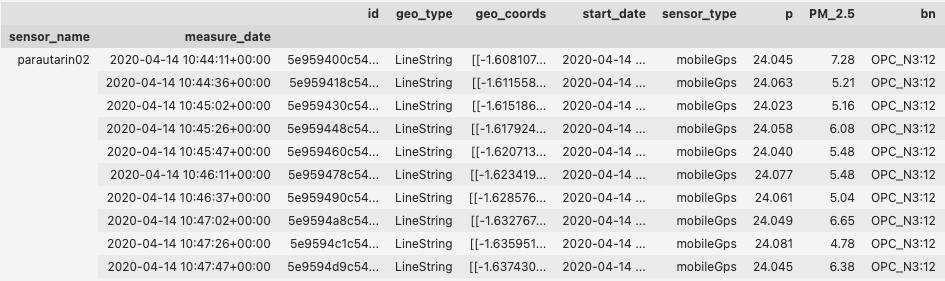}
    \caption{Sample of the small AQMO dataset - not all columns are shown}
    \label{fig:small-dataset}
\end{figure}

The aim of this smaller dataset is to access easily to features; in order to produce statistics. It is distributed in several formats : .csv, .pkl\footnote{a python pickle containing \href{https://pandas.pydata.org/pandas-docs/stable/reference/api/pandas.DataFrame.html}{pandas DataFrame} - note that pickles should be used carefully because they may contain arbitrary code}. For fast data analysis, this dataset should be preferred rather than the raw one described above. To get access to deeper information, one can still load the raw dataset. See Figure \ref{fig:small-dataset} for an overview.

Several choices were made in order to get a more digest dataset :
\begin{itemize}
    \item Rows with missing values in at least one of the columns : 'PM\_2.5', 'p' (time precision) and 'bn' (Base Name of the sensor) were removed.
    \item Duplicated measures were aggregated to the mean of the most precise ones. Some values were delayed and therefore the known measure date is inexact and some measure dates were found to have several values (up to 6000 !). To assign each date-time a single measure, we made the following computation : for measure date with more than one measure, we computed the PM\(_{2.5}\) value for this date by taking the average among the measures with minimal time precision ('p').
    \item Longitude and Latitude values for geojson LineString were set to the means along all longitudes and latitudes of the LineString.
\end{itemize}
The extraction was made using basic Python libraries like \href{https://pandas.pydata.org/}{pandas}. Here is a python one-liner to load the dataset in a pandas DataFrame :

\begin{lstlisting}[language=Python]
df = pd.read_csv(
        "https://opendata.fenix.rudi-univ-rennes1.fr/storage/download/"
        "97f3614a-6d0f-47c5-bc96-eb9a0c789d27",
        compression="zip",
        index_col=[0, 1]
    )
\end{lstlisting}

We nevertheless advise you to download the dataset locally once and for all :
\begin{itemize}
    \item link to the .csv : \url{https://opendata.fenix.rudi-univ-rennes1.fr/storage/download/9f57dedc-f6c7-4172-a79c-5ce7f104bc88};
    \item link to the .pkl : \url{https://opendata.fenix.rudi-univ-rennes1.fr/storage/download/e78b51a1-b8f8-4932-bd98-1198b2897353};
    \item link to the scheme :
    \url{https://opendata.fenix.rudi-univ-rennes1.fr/storage/download/d07b7eeb-0994-4d0f-ac03-67b0d20178bd};
    \item link to the notebook describing all steps of the transformation from raw data to small tabular dataset : \url{https://github.com/MatthieuAdafaly/AQMO/blob/master/Data_Import.ipynb}.
    \item more detailed metadata about the tabular dataset on the RUDI-node metadata catalog of the University of Rennes : \url{https://opendata.fenix.rudi-univ-rennes1.fr/catalog/v1/resources/9e64890a-7d31-4ad8-8011-356e0ef2fceb}.
\end{itemize}

\begin{figure}[!ht]
    \centering
    \includegraphics[width=0.8\textwidth]{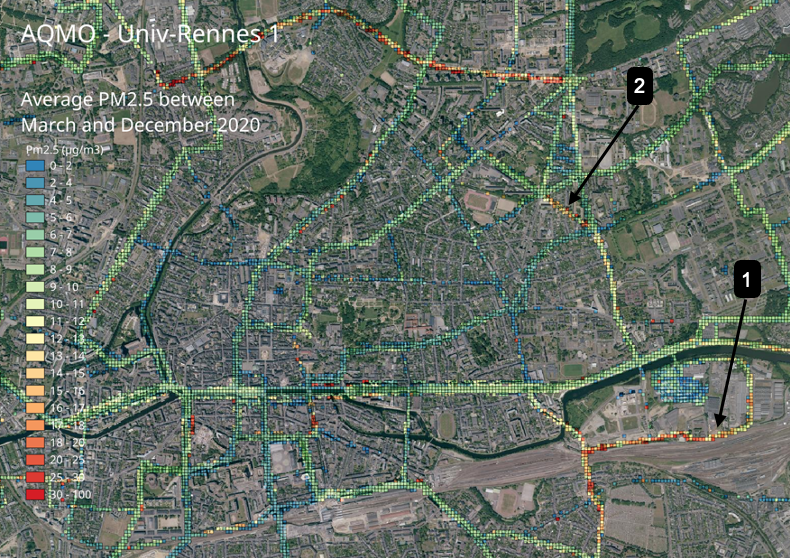}
    \caption{Example of data use to identify points of interest: one near the railway technical center and another on a specific street.}
    \label{fig:aqmo_points_interest}
\end{figure}

\section{Potential Reuse Example}

Illustrated in Figure~\ref{fig:aqmo_points_interest}, the sensors have already demonstrated their ability to detect significant transient PM\(_{2.5}\) pollution variations across the city. Among these, at least two unexpected points of interest have been identified: a specific street in the city and an industrial area near the railway technical center. 

\section*{Acknowledgements}
We thank Rennes Metropolis and Keolis for their help with the deployment of the microsensors.

\section*{Funding}
AQMO has been funded by CEF (Grant Agreement Number: INEA/CEF/ICT/A2017/1566962).
The Connecting Europe Facility (CEF) is a European Union funding instrument aimed at promoting growth, employment, and competitiveness through targeted investments in pan-European infrastructure (\url{https://commission.europa.eu/funding-tenders/find-funding/eu-funding-programmes/connecting-europe-facility_en}).

\bibliographystyle{johd}
\bibliography{references}
\end{document}